\documentclass[12pt]{iopart}
\usepackage{amssymb} 
\usepackage{braket} 
\usepackage{graphicx} 
\usepackage{subfigure}
\usepackage{booktabs}
\usepackage{bbm}

\begin{document}

\title{Cooperative scattering of scalar waves by optimized configurations of point scatterers}

\author{Frank Sch\"afer, Felix Eckert and Thomas Wellens}

\address{Physikalisches Institut, Albert-Ludwigs-Universit\"at Freiburg, Hermann-Herder-Str. 3, D-79104 Freiburg, Germany}
\eads{\mailto{frank.schaefer@uranus.uni-freiburg.de} and \mailto{thomas.wellens@physik.uni-freiburg.de}} 
\vspace{10pt}
\begin{indented}
\item[]\today
\end{indented}

\begin{abstract}

We investigate multiple scattering of scalar waves by an ensemble of $N$ resonant point scatterers in three dimensions.
For up to $N=21$ scatterers, we numerically optimize the positions of the individual scatterers, such as to maximize the total scattering cross section for an incoming plane wave, on the one hand, and to minimize the decay rate associated to a long-lived scattering resonance, on the other hand. In both cases, the 
optimimum is achieved by configurations where all scatterers are placed on a line parallel to the direction of the incoming plane wave.
The associated maximal scattering cross section increases quadratically with the number of scatterers for large $N$, whereas the minimal decay rate  -- which is realized by configurations that are not the same as those that maximize the scattering cross section -- decreases exponentially as a function of $N$. Finally, we also analyze the stability of our  optimized configurations with respect to small random displacements of the scatterers. These results demonstrate that optimized configurations of scatterers bear a considerable potential for applications such as quantum memories or mirrors consisting of only a few atoms.

 \end{abstract}

%
\vspace{2pc}
\maketitle
%
%

\section{Introduction}

In general, a collection of $N$ point scatterers, with distances between the scatterers of the order of the wavelength, scatters an incoming coherent wave in a different way than $N$ independent scatterers which are placed far away from each other. 
An obvious example of such \lq cooperative scattering\rq\ is observed in a periodic crystal of scatterers, where the fields emitted by each individual scatterer interfere contructively with each other in certain directions and destructively in other directions (Bragg scattering \cite{Bragg}). Recently, Bragg scattering has been exploited to realize strong reflection of light from only one or two thousand atoms aligned along an optical fiber \cite{Corzo,Sorensen}.

Apart from interference between singly scattered waves, cooperative effects are also induced by multiple scattering, where waves emitted by individual scatterers are again scattered by other scatterers, and so on. If the wave is scattered back and forth many times between a few scatterers, this may lead to a strong enhancement of the local field intensity in the vicinity of these scatterers. These local field enhancements are interesting for technological applications, since they can be used to increase the efficiency of solar cells or other optical devices \cite{hofmann,spallek}, and have been observed in specifically tailored nanostructured materials consisting of, e.g. metallic nanoantennas \cite{podolskyi}, nanospheres \cite{Li} or nanoparticles \cite{Wang}. 

In the present article, we will investigate scalar point scatterers as a paradigmatic model system for multiple scattering.
A single point scatterer exhibits a maximum scattering cross section of $\sigma_{\rm max}^{(1)}=\lambda^2/\pi$, where $\lambda$ denotes the wavelength of the scattered light \cite{niuwenhuizen}. 
The fundamental questions which we address in the present article are: what is the largest scattering cross section that can be achieved with an ensemble of $N$ point scatterers, and how must the scatterers be positioned in space in order to realize this maximum? In particular, we will show that, by choosing optimized configurations of point scatterers, it is possible to exploit effects of cooperative scattering such that the total scattering cross section exceeds the value $N\sigma_{\rm max}^{(1)}$ for $N$ independent scatterers. Expressing the total scattering cross section as a sum over scattering resonances, we furthermore demonstrate that a necessary (but not sufficient) condition to achieve a large scattering  cross section is the existence of a narrow scattering  resonance with correspondingly small decay rate.  In quantum optics, such states are known as \lq subradiant states\rq\ or \lq dark states\rq\ \cite{pavolini}, and have recently been observed in a large cloud of cold atoms \cite{guerin}.

The article is structured as follows: After briefly outlining the theoretical frame (see chapter \ref{sec:theory}), we will 
describe the numerical optimization algorithm and present our results for optimized configurations which maximize the total  scattering cross section, on the one hand, and those which minimize the smallest decay rate, on the other hand (chapter \ref{sec:results}). Furthermore, we will analyze the stability of our optimized configurations with respect to small random displacements of the scatterers (chapter \ref{sec:stability}). 
The last section concludes this article and gives a short outlook on future perspectives.

\section{Theoretical frame}
\label{sec:theory}

In this section, we review basic theoretical concepts of scattering from a collection of point scatterers.
After introducing fundamental equations describing scattering of a monochromatic, scalar wave by a single point scatterer, we  treat the case of many point scatterers based on the Foldy-Lax formalism for multiple scattering \cite{foldy,lax}. 
We  derive  
the total scattering cross section for $N$ scatterers and express it in terms of scattering resonances, where the widths of these resonances correspond to the decay rates of the corresponding resonance states.

\subsection{Scalar wave equation}

We consider the following time-independent equation for a scalar wave $\Psi({\bf r})$ in three dimensional real space:
\begin{equation}
\vec{\nabla}^2\Psi({\bf r})+k^2 [1+\chi({\bf r})]\Psi({\bf r}) =0\label{eq:wave}
\end{equation}
with wavenumber $k$ and a spatial inhomogeneity $\chi({\bf r})$ that induces scattering. For the case of a quantum particle obeying the Schr\"odinger equation, for example, the spatial inhomogeneity is given by $\chi({\bf r})=-2mV({\bf r})/(\hbar^2 k^2)$ in terms of the scattering potential $V({\bf r})$ and the mass $m$. Solutions of the corresponding time-dependent  wave equation are given by 
\begin{equation}
\tilde{\Psi}({\bf r},t)=\Psi({\bf r}) e^{-i\omega t}\label{eq:time}
\end{equation}
where the frequency $\omega$ is determined by $k$ through the dispersion relation (e.g. $\omega=\frac{\hbar k^2}{2m}$ in case of the Schr\"odinger equation). In the following, however, we will restrict ourselves to the stationary regime described by Eq.~(\ref{eq:wave}).

Electromagnetic waves propagating 
in dielectric media with space-dependent electric susceptibility $\chi({\bf r})$ are also described by
Eq.~(\ref{eq:wave}) in a scalar approximation (i.e. ignoring the vectorial character of the electric field) which is often used in the theory of multiple scattering of light (e.g. \cite{rossum}). Nevertheless, scalar waves and vectorial waves may behave differently, especially if the distances between scatterers are smalll \cite{skipetrov}. Therefore, the optimal configurations 
of point scatterers investigated in this article apply, in principle, only to scalar waves, although we expect the optimal configurations for vector waves to have similar qualitative properties. In the vectorial case, cooperative effects of light scattering in one-dimensional atomic arrays have been studied recently \cite{bettles}, but only for periodic arrays (i.e. without optimizing the individual positions of the scatterers).

\subsection{Single point scatterer}

A point scatterer at position ${\bf r}_0$ is defined by the condition that 
the inhomogeneity $\chi({\bf r})$ in Eq.~(\ref{eq:wave}) is different from zero only in a small volume containing the point ${\bf r}_0$, i.e.  if $|{\bf r}-{\bf r}_0|\ll 1/k$. The solution of Eq.~(\ref{eq:wave}) with incoming plane wave 
\begin{equation}
\Psi_0({\bf r})=\Psi_0 e^{i{\bf k}_{\rm in}\cdot{\bf r}}
\end{equation}
where ${\bf k}_{\rm in}$ with $|{\bf k}_{\rm in}|=k$ denotes the incoming wave vector, 
 is then given by \cite{heller1}:
\begin{eqnarray}
\Psi(\bi{r})&=\Psi_0({\bf r})+\frac{e^{ik|\bi{r}-\bi{r}_0|}}{|\bi{r}-\bi{r}_0|}f\Psi_0({\bf r}_0)
\label{eq:wavefunctionsinglepoint}
\end{eqnarray}
where $e^{ik|\bi{r}-\bi{r}_0|}/(|\bi{r}-\bi{r}_0|)$
 represents a spherical wave at $\bi{r}$ originating from point $\bi{r}_0$, and $f$ denotes the scattering amplitude.
 The latter is isotropically distributed (i.e. exhibits no angular dependence) 
 due to the  negligible extent of the scatterer. In principle, the scattering amplitude $f$ can be calculated for a given $\chi({\bf r})$ through the solution of the Lippmann-Schwinger equation \cite{lippmann}. In the following, however, we will not be concerned with the actual physical realization of the point scatterer, but  rather assume that its properties are given to us in terms of its scattering amplitude $f$.

 The scattering cross section of a single point scatterer is given by:
\begin{equation}
\sigma^{(1)}=4\pi |f|^2
\end{equation}
 Moreover, the scattering amplitude obeys the optical theorem which guarantees flux conservation \cite{rodberg}:
\begin{eqnarray} &k |f|^2 ={\rm Im}\left[f\right]. 
\label{eq:opticaltheorem} 
\end{eqnarray}
The general solution of equation (\ref{eq:opticaltheorem}) (for $f\neq 0$) can be written in the following form:
\begin{eqnarray}
f=\frac{1}{k(\alpha -i)}, \quad \alpha \in \mathbb{R}.
\label{eq:scatteringamplitude}
\end{eqnarray}
The optical theorem thus yields a fundamental upper limit for the scattering cross section of a point scatterer, which is independent of its physical realization. This maximum is achieved by choosing $\alpha=0$ in (\ref{eq:scatteringamplitude}), leading to:
\begin{eqnarray}
\sigma_{\rm max}^{(1)}=\frac{4\pi}{k^2}.
\end{eqnarray}
In general, the real parameter $\alpha$ determining the scattering amplitude (\ref{eq:scatteringamplitude})  depends on the
wavenumber $k$ or, equivalently (due to the dispersion relation),  on the frequency $\omega$  of the scattered wave. The condition $\alpha(\omega_0)=0$ then corresponds to a resonance condition with resonance frequency $\omega_0$.
For small deviations $\delta=\omega-\omega_0$ from the resonance, we can expand $\alpha$ in first order around $\omega_0$, such that
\begin{equation}
f=\frac{1}{k\left(\frac{\delta}{\gamma} -i\right)}\label{eq:fdelta}
\end{equation}
where $\gamma$ defines the width of the resonance. 

\subsection{Many point scatterers}
For $N$ point scatterers placed at positions ${\bf r}_1,\dots,{\bf r}_N$, the following set of equations
for the fields $\Psi_i$ incident on scatterer $i$ have been derived   \cite{foldy,lax}:
\begin{equation}
\Psi_i=\Psi_0({\bf r}_i)+k\sum_{j=1}^{N}G_{ij}f_j \Psi_j
\label{eq:fieldatscatterer}
\end{equation}
where $f_i$ denotes the scattering amplitude of scatterer $i$, and the Green matrix $G$ is defined as
\begin{equation}
G_{ij}=\cases{0&  if  $i = j$\\ \frac{e^{ik|{\bf r}_i-{\bf r}_j|}}{k|{\bf r}_i-{\bf r}_j|}&  if $i \neq j$\\}
\label{eq:Greensmatrix}
\end{equation} 
According to (\ref{eq:fieldatscatterer}) and (\ref{eq:Greensmatrix}), the field at scatterer $i$ is obtained as the incoming wave $\Psi_0({\bf r}_i)$ plus the fields scattered from all other scatterers $j\neq i$. Similarly,
the total wave function $\Psi({\bf r})$ (at any position ${\bf r}$ different from ${\bf r}_i$, $i=1,\dots,N$) is obtained as:
\begin{equation}
\Psi({\bf r})=\Psi_0({\bf r})+\sum_{i=1}^{N}\frac{e^{ik|{\bf r}-{\bf r}_i|}}{|{\bf r}-{\bf r}_i|} f_i\Psi_{i}
\end{equation}
In order to determine the scattering cross section $\sigma$, we calculate the flux of the scattered part $\Psi_{\rm sc}({\bf r}):=
\Psi({\bf R})-\Psi_0({\bf r})$ of the
wave through a sphere with large radius $R$. By definition, this flux is equal to the flux of the incoming wave through the area $\sigma$.
Applying a far-field approximation ($|{\bf R}|\gg |{\bf r}_i|$, $i=1,\dots,N$), we obtain:
\begin{equation}
\Psi_{\rm sc}({\bf R})=\frac{e^{i k R}}{R}\sum_{i=1}^{N} e^{-i{\bf k}_{\rm out}\cdot{\bf r}_i}  f_i\Psi_{i}\label{eq:Psisc}
\end{equation}
where ${\bf k}_{\rm out}=k {\bf R}/|{\bf R}|$ denotes the wavevector of the outgoing wave. Since, in the far field, the flux density
${\bf J}_{\rm sc}=-\frac{1}{k}\frac{{\rm d}\omega}{{\rm d}k} {\rm Im}[\Psi_{\rm sc}\vec{\nabla}\Psi_{\rm sc}^*]$ 
points in the radial direction, the total scattering cross section is obtained by integrating the wave intensity (normalized by the incoming intensity) over the surface of a sphere with radius $R$:
\begin{equation}
\sigma=\int{\rm d}\Omega~R^2 \frac{|\Psi_{\rm sc}({\bf R})|^2}{|\Psi_0|^2}=\int{\rm d}\Omega 
\left|\sum_{i=1}^N e^{-i{\bf k}_{\rm out}\cdot{\bf r}_i} f_i\frac{\Psi_i}{\Psi_0}\right|^2
\label{eq:crosssecangle}
\end{equation}
where $\Omega$ represents the angular variables defining the direction of ${\bf R}$ (and of ${\bf k}_{\rm out}$).
Using
\begin{equation}
\int{\rm d}\Omega~e^{-i{\bf k}_{\rm out}\cdot ({\bf r}_i-{\bf r}_j)} = 4\pi \left(\frac{G_{ij}-G^*_{ij}}{2i}+\delta_{ij}\right)
\end{equation}
see equation (\ref{eq:OmG}), we obtain:
\begin{eqnarray}
\sigma & = & 4\pi \sum_{i,j=1}^N \frac{f_i \Psi_i f_j^*\Psi_j^*}{|\Psi_0|^2}  \left(\frac{G_{ij}-G^*_{ij}}{2i}+\delta_{ij}\right)\nonumber\\
& = & \frac{4\pi}{k|\Psi_0|^2} \sum_{i=1}^N {\rm Im}\left[f_i \Psi_i \Psi_0^*({\bf r}_i)\right]\label{eq:crosssectionN}
\end{eqnarray}
where we used (\ref{eq:opticaltheorem}) and (\ref{eq:fieldatscatterer}) in the second line. Equation (\ref{eq:crosssectionN}) expresses the optical theorem for the $N$-scatterer system: the total scattering cross section is associated with destructive interference of the waves emitted by scatterers $i=1,\dots,N$ with the incoming wave, or, in simpler words, due to flux conservation, the scattered flux is taken away from the incoming flux.

\subsection{Scattering resonances}
\label{sec:resonances}

From now on, we assume that all $N$ scatterers are identical, i.e. $f_i=f$ for all $i=1,\dots,N$.
Defining $N$-dimensional vectors $|{\bf \Psi}\rangle$ and $|{\bf \Psi_0}\rangle$ with components $\langle i|{\bf \Psi}\rangle:=\Psi_i$ and
$\langle i|{\bf \Psi_0}\rangle:=\Psi_0({\bf r}_i)$, a formal solution of (\ref{eq:fieldatscatterer}) is obtained as follows:
\begin{equation}
|\Psi\rangle = ({\mathbbm 1}-k f G)^{-1} |\Psi_0\rangle
\end{equation}
Using the explicit form (\ref{eq:fdelta})  of the scattering amplitude, we can rewrite the scattering cross section (\ref{eq:crosssectionN}) as a function of the detuning $\delta$ as follows:
\begin{equation}
\sigma(\delta)=\frac{4\pi}{k^2|\Psi_0|^2}{\rm Im}\left[\left<{\bf \Psi_0}\left|\left(\frac{\delta}{\gamma} -i - G\right)^{-1}\right|{\bf \Psi_0}\right>\right]\label{eq:crosssectionvec}
\end{equation}
Since $\omega=\omega_0+\delta$ is related to $k$ through the dispersion relation, we must be aware of the fact that 
$k$ (and, consequently also $G$) depends on $\delta$. These dependencies can be neglected, however,
if we assume  $|\delta|\ll k\left|\frac{{\rm d}\omega}{{\rm d}k}\right|$ and $|\delta|\ll \left|\frac{{\rm d}\omega}{{\rm d}k}\right|/r_{ij}$
for all distances $r_{ij}$. This, in turn, is justified in the case of a narrow single-scatterer resonance (i.e. if 
$\gamma\ll k\left|\frac{{\rm d}\omega}{{\rm d}k}\right|$ and $\gamma\ll \left|\frac{{\rm d}\omega}{{\rm d}k}\right|/r_{ij}$), since the 
scattering cross section is negligibly small for $|\delta|\gg\gamma$.

From (\ref{eq:crosssectionvec}), we see that each eigenvalue $\lambda_n$ of $G$, with corresponding right-eigenvector
$|{\bf \lambda_n^{(R)}}\rangle$ and left-eigenvector $\langle{\bf \lambda_n^{(L)}}|$ normalized such that
$\langle{\bf \lambda_n^{(L)}}|{\bf \lambda_n^{(R)}}\rangle=1$, gives rise to a scattering resonance:
\begin{equation}
\sigma_n(\delta)=\frac{4\pi\gamma}{k^2|\Psi_0|^2}{\rm Im}\left[\frac{\langle{\bf \Psi_0}|{\bf \lambda_n^{(R)}}\rangle\langle{\bf \lambda_n^{(L)}}|{\bf \Psi_0}\rangle}{\delta-\delta_n-i\gamma_n}\right]\label{eq:resonancen}
\end{equation}
where the position $\delta_n$ and the width $\gamma_n$ are determined by the real and imaginary part of the eigenvalue $\lambda_n$ as follows:
\begin{eqnarray}
\delta_n & = & \gamma {\rm Re}[\lambda_n] \\
\gamma_n & = & \gamma\left(1+{\rm Im}[\lambda_n]\right)\label{eq:gamman}
\end{eqnarray}
If the matrix $G$ is diagonalisable -- which is the case for all the optimized configurations discussed below (see also the last paragraph of \ref{sec:appendix}) -- the total scattering cross section results as the sum over all $N$ resonances:
\begin{equation}
\sigma(\delta)=\sum_{n=1}^N \sigma_n(\delta)
\label{eq:sigman}
\end{equation}
Since, in general, the left- and right-eigenvectors differ from each other (see \ref{sec:appendix}), the factor $\langle{\bf \Psi_0}|{\bf \lambda_n^{(R)}}\rangle\langle{\bf \lambda_n^{(L)}}|{\bf \Psi_0}\rangle$ appearing in Eq.~(\ref{eq:resonancen}) may exhibit a non-vanishing imaginary part leading to an asymmetric, Fano-like profile of the scattering resonance $\sigma_n(\delta)$ \cite{bettles,fano}.

\subsection{Decay rates}
\label{sec:decayrates}

Setting $\Psi_i=\langle i|{\bf \lambda_n^{(R)}}\rangle$ and choosing a complex frequency 
$\omega=\omega_0+\delta_n+i\gamma_n$ [i.e. $\delta=\delta_n+i \gamma_n$ in equation~(\ref{eq:fdelta}), while neglecting the frequency dependence of $k$ and $G$, as mentioned above], we obtain a solution of the multiple scattering equation (\ref{eq:fieldatscatterer})   without incoming wave, i.e. for $\Psi_0({\bf r}_i)=0$. Since an imaginary part of $\omega$ leads to an exponential decay in time, see equation (\ref{eq:time}), the resonance width $\gamma_n$ yields the decay rate of the corresponding eigenstate $|{\bf \lambda_n^{(R)}}\rangle$. Of special interest are states which exhibit very small decay rates: if such a state is prepared at time $t=0$, it will be stored in the system for a very long time. The physical reason for the long lifetime is that the spherical waves emitted by the individual scatterers interfere destructively with each other.

Such long-lived, subradiant states arise naturally if two scatterers are placed very close to each other \cite{heller2}. 
From a practical point of view, however, it is not possible to choose the distance arbitrarily small, since every physical scatterer has a finite size. We will therefore show below that strongly subradiant states can also be constructed with scatterers that are not very close together.

Choosing $\delta=\delta_n$ in (\ref{eq:resonancen}), where $\gamma_n$ appears in the denominator, one could expect that states with very small decay rate $\gamma_n$ also lead to a very large scattering cross section. As it turns out, however, the enumerator $\langle{\bf \Psi_0}|{\bf \lambda_n^{(R)}}\rangle\langle{\bf \lambda_n^{(L)}}|{\bf \Psi_0}\rangle$, which describes the overlap of the resonance state with the incoming wave, also tends to zero if $\gamma_n\to 0$ (see \ref{sec:appendix}). This can be understood in terms of a reciprocity argument: states which do not emit outgoing waves are effectively isolated from the rest of the system and cannot be excited by any incoming wave. Maximizing the scattering cross section therefore amounts to a compromise between minimizing the decay rate and maximizing the overlap with the incoming wave.
As shown in \ref{sec:appendix}, this compromise amounts to concentrating the angular profile of the scattered wave into the forward and the backward direction, whereas keeping scattering into all other directions as small as possible.

\section{Optimized configurations}
\label{sec:results}

\begin{figure}\begin{center}\includegraphics[width=0.9\textwidth]{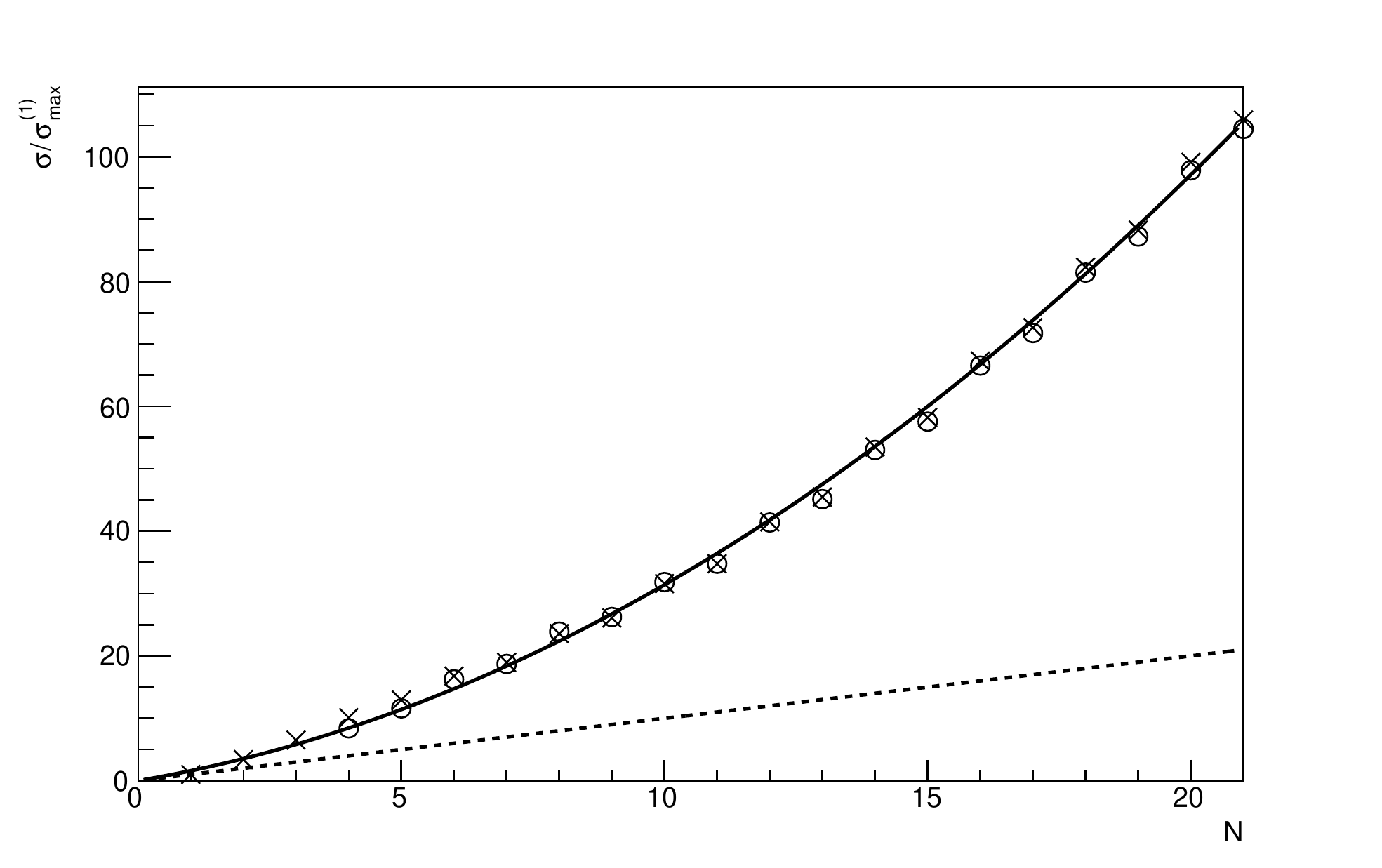}
\caption{Optimized scattering cross sections $\sigma$ in units of the single-scatterer cross section $\sigma^{(1)}_{\rm max}=4\pi/k^2$ as 
a function of the number $N$ of resonant ($\delta=0$) point scatterers. Crosses (circles) refer to the optimized narrow (wide) configurations, see figure~\ref{fig:config_eng} (figure~\ref{fig:config_breit}). The fit function $\sigma(N)=a N^2+b N$ (solid line),
with $a=0.172 \pm 0.005$ and $b=1.43 \pm 0.09$ as fitting parameters,
 shows that the optimized scattering cross sections grow quadratically as a function of $N$ for large $N$.
For comparison, the dotted line displays the result $\sigma=N\sigma^{(1)}_{\rm max}$ obtained for independent scatterers. 
\label{fig:crosssec}}
\end{center}
\end{figure}

In this section, we present our numerical results for the maximization of the scattering cross section, as well as for the minimization of the smallest decay rate among the $N$ eigenstates of the Green matrix. In both cases, we look for optimal configurations by varying the positions of the $N$ scatterers.

\subsection{Numerical algorithm}
\label{sec:algorithm}

Our algorithm for the optimization of the positions of the scatterers in order to maximize the scattering cross section  (\ref{eq:crosssectionvec}) is constructed as follows \cite{frank}: The scatterers are randomly distributed in a sphere with a radius $R$. At least 10 000 random configurations are chosen, and the scattering cross section is evaluated for each of them. For the best of these random positions, a downhill simplex-algorithm \cite{c} searches for the local maximum. These steps are repeated at least 1000 times for each value of the radius $R$, and, finally, the whole procedure is repeated again for 
different radii {\bf ($kR=1,2,\dots,12$)}. At the end, we save the ten best configurations found in this way with the associated optimized scattering cross sections. For simplicity, we restrict the optimization  to the case $\delta=0$, where the frequency of the incoming wave exactly coincides with the resonance frequency of the individual scatterers. As we have checked, an additional optimization of the incident frequency leads only to a small improvement of the scattering cross section (of the order of a few percent for odd numbers of scatterers, and even less for even numbers).

The program for the minimization of the decay rate  \cite{felix} is designed in a similar manner, with the exception that the minimization is performed under the constraint of a minimal exclusion radius $r_{\rm excl}$ around each scatterer to ensure that the scatterers are not at the same place, see the discussion in chapter~\ref{sec:decayrates}.

\subsection{Maximization of the scattering cross section}
\label{sec:maxcrosssec}

\begin{figure}\begin{center}\includegraphics[width=0.9\textwidth]{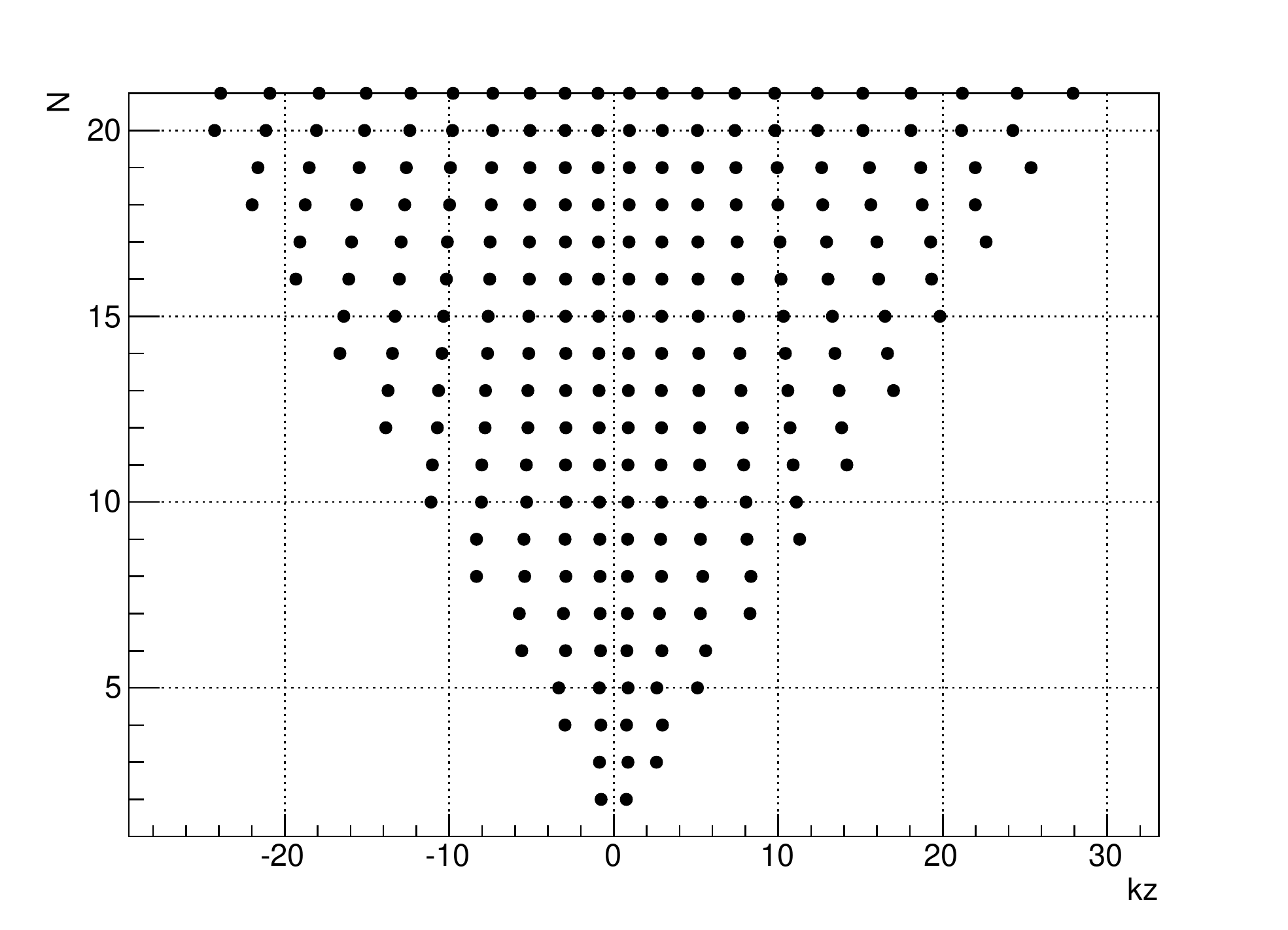}
\caption{Optimal \lq narrow\rq\  configurations (without gap in the center, cf. Fig.~\ref{fig:config_breit}) maximizing the scattering cross section for different  numbers $N=2,\dots,21$ of resonant point scatterers (i.e. $\delta=0$). All scatterers are placed on a line parallel to the direction $z$ of the incoming wave.
Configurations with even $N$ are mirror-symmetric with respect to $z=0$.
\label{fig:config_eng}}
\end{center}
\end{figure}

\begin{figure}\begin{center}\includegraphics[width=0.9\textwidth]{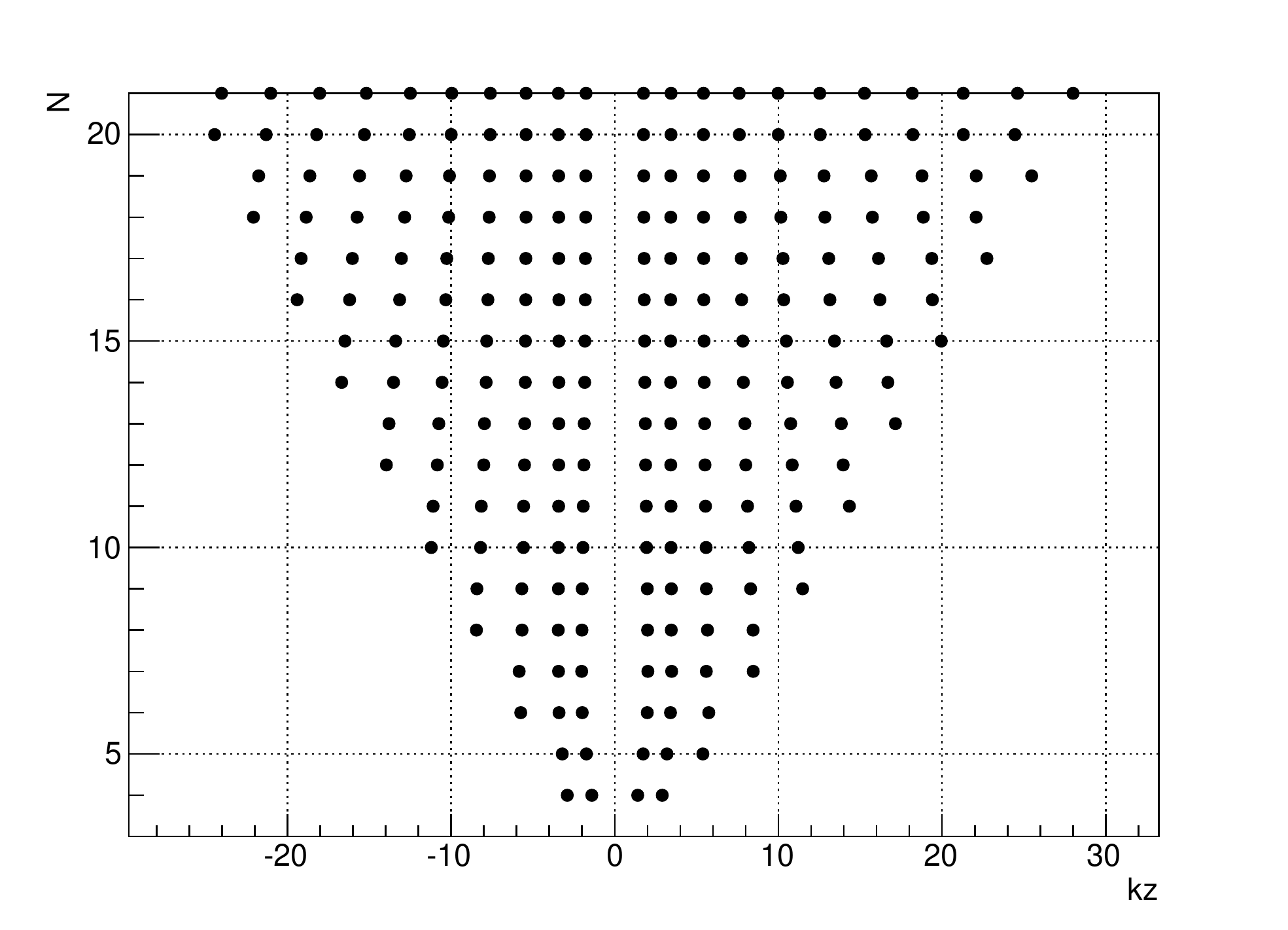}
\caption{Optimal \lq wide\rq\ configurations (exhibiting a gap in the center) maximizing the scattering cross section for different numbers 
$N=4,\dots,21$ of resonant point scatterers (i.e. $\delta=0$). As already observed for the narrow configurations, see Fig.~\ref{fig:config_eng}, all scatterers are placed on a line parallel to the direction $z$ of the incoming wave, and
configurations with even $N$ are mirror-symmetric with respect to $z=0$. The wide configurations exhibit almost the same scattering cross section as  the narrow configurations, see Fig.~\ref{fig:crosssec}. \label{fig:config_breit}}
\end{center}
\end{figure}

For $N\leq 6$ scatterers, the optimal configurations found by the above algorithm exhibit the property that all
scatterers are placed on a line parallel to the direction of the incoming plane wave. For $N>6$, it becomes increasingly difficult and time-consuming to find the optimal configuration in the $3N$-dimensional space of position variables. In this case, better results are obtained if the optimization is restricted to one-dimensional configurations in accordance with the 
finding for $N\leq 6$.
For even $N$, it turns out to be advantageous to further restrict the search to symmetric configurations (see below). For large 
numbers of scatterers ($N>18$ for even $N$, and $N>10$ for odd $N$), we obtain the largest scattering cross sections by starting with optimized configuration for $N-2$, placing two additional scatterers at both ends of the chain, and applying the downhill simplex algorithm to this configuration.

Since an analytical upper bound of the scattering cross section is not available (see also \ref{sec:appendix}), there is no strict proof that our numerically optimized configurations are truly the optimal ones, although we  suspect that this is the case. Apart from numerical evidence, 
another indication that our configurations are indeed the optimal ones is the fact that they behave in a regular way as a function of $N$ (see figures \ref{fig:config_eng} and \ref{fig:config_breit}).

The corresponding optimized scattering cross sections are shown in figure~\ref{fig:crosssec} as a function of $N$. For each value of $N\geq 4$, we find two different optimized configurations (a \lq narrow\rq\ one and a \lq wide\rq\ one, see figures~\ref{fig:config_eng} and \ref{fig:config_breit}) which exhibit almost the same scattering cross section. For $N=8$ and $N=10$ the scattering cross section of the  wide configuration (circles) is slightly larger than the scattering cross section of the narrow configuration (crosses), and vice versa for all other values of $N$ up to $N=21$. The fit function $\sigma(N)=a N^2+b N$ (solid line) indicates that the optimized scattering cross sections grow quadratically as a function of $N$ for large $N$. They hence  exceed  the linearly increasing result  obtained for independent scatterers that are placed far away from each other (dotted line). This is remarkable since, as we have verified, the scattering cross sections of non-optimized, random configurations of scatterers (with distances between the scatterers of the order of the wavelength or smaller) are typically much smaller than $N\sigma^{(1)}_{\rm max}$. For one-dimensional, regular arrays (with lattice spacing $a$ between neighbouring scatterers), we have checked that the scattering cross section $\sigma(N)$ grows linearly with $N$ for large $N$ (and only slightly exceeds the value
$N\sigma^{(1)}_{\rm max}$ for independent scatterers), even if the lattice spacing $a$ is optimized for each value of $N$.
Therefore, only by choosing the positions of all  scatterers in an optimal way, it is possible to induce cooperative effects of multiple scattering that increase the scattering cross section significantly beyond the value of $N$ independent scatterers.

The corresponding optimal positions are shown in figures~\ref{fig:config_eng} and \ref{fig:config_breit} for the narrow and wide configurations, respectively. As already mentioned above, all scatterers are arranged on a line parallel to the direction ${\bf k}_{\rm in}$ of the incoming wave (here chosen as the $z$-direction). For even numbers of scatterers, we find that the optimized positions (in both configurations) are mirror-symmetric with respect to a certain symmetry point, which we choose as $z=0$ (since the cross section is invariant under constant translations of all scatterers).
For odd numbers of scatterers, the configurations are similar to the symmetric ones for $N-1$ scatterers, with one additional scatterer placed at one side of the chain. (We choose the right-hand side in figures~\ref{fig:config_eng} and \ref{fig:config_breit}, but, due to the symmetry of the scattering cross section with respect to inverting the direction of the incoming wave, ${\bf k}_{\rm in}\to -{\bf k}_{\rm in}$, see \ref{sec:appendix}, the same scattering cross section is obtained when placing the additional scatterer at the left-hand side. Moreover, the point $z=0$ is choosen such that
$z_{(N-1)/2}=-z_{(N+1)/2}$, see also table~\ref{tab:appendixB}.)

\begin{figure} 
	\begin{center}
		\includegraphics[width=0.9\textwidth]{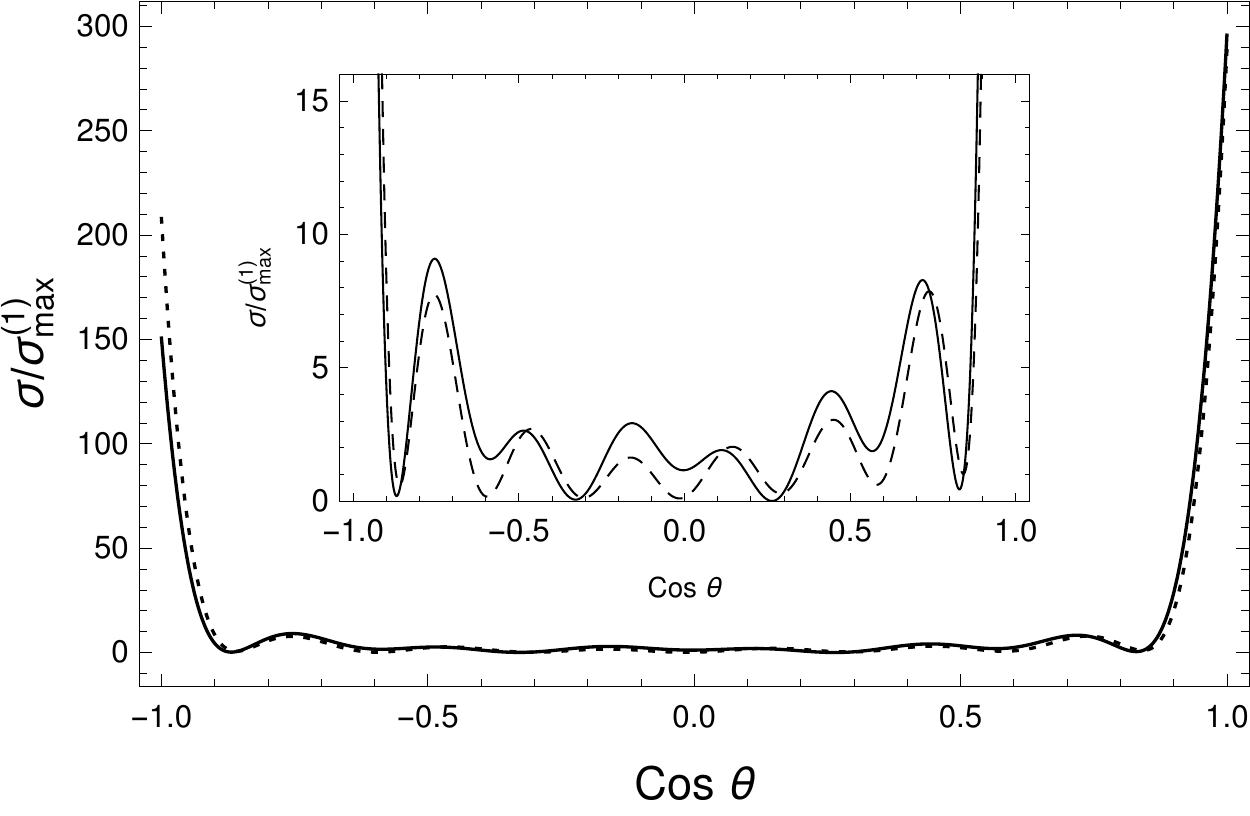}
	\end{center}
	\caption{Angular distribution $\sigma(\cos\theta)$ of the scattered wave (in units of the single-scatterer cross section $\sigma^{(1)}_{\rm max}=4\pi/k^2$) for the optimal narrow configuration (dashed line) and the  optimal wide configuration (solid line) with $N=8$ scatterers. Both angular distributions exhibit pronounced emissions into the forward and backward direction ($\cos\theta=+1$ and $\cos\theta=-1$, respectively), while keeping the emissions into all other directions relatively small. The inset shows a zoom for small values of  $\sigma(\cos\theta)$. \label{fig:angular_distribution}}	
\end{figure}

As shown in \ref{sec:appendix}, the maximization of the scattering cross section amounts to concentrating the angular profile of the scattered wave into the forward and backward directions (${\bf k}_{\rm out}=\pm{\bf k}_{\rm in}$), while keeping the emission into other directions as small as possible. 
In figure~\ref{fig:angular_distribution}, we therefore plot the angular profile $\sigma(\cos\theta)$ of the scattering cross section, defined by equation (\ref{eq:crosssecangle}) without integral over $\Omega$,
i.e. $\sigma(\cos\theta)=2\pi \left|\sum_i e^{ikz_i\cos\theta} f_i\Psi_i/\Psi_0\right|^2$ (where we used the fact that the optimized configurations are one-dimensional, and therefore the differential cross section is independent of the azimuthal angle $\phi$).
Indeed, we see two pronounced peaks in the forward and backward directions, $\cos\theta=+1$ and $\cos\theta=-1$, respectively.
According to this result, it is not surprising that the optimal configurations are one-dimensional (with scatterers placed on a line parallel to ${\bf k}_{\rm in}$), since, for such a configuration, it is intuitively plausible that a large asymmetry between the parallel and perpendicular directions can be achieved. 

\subsection{Scattering resonances of optimized configurations}
\label{sec:eigenmodes}

\begin{figure} 
	\begin{center}
		\includegraphics[width=0.9\textwidth]{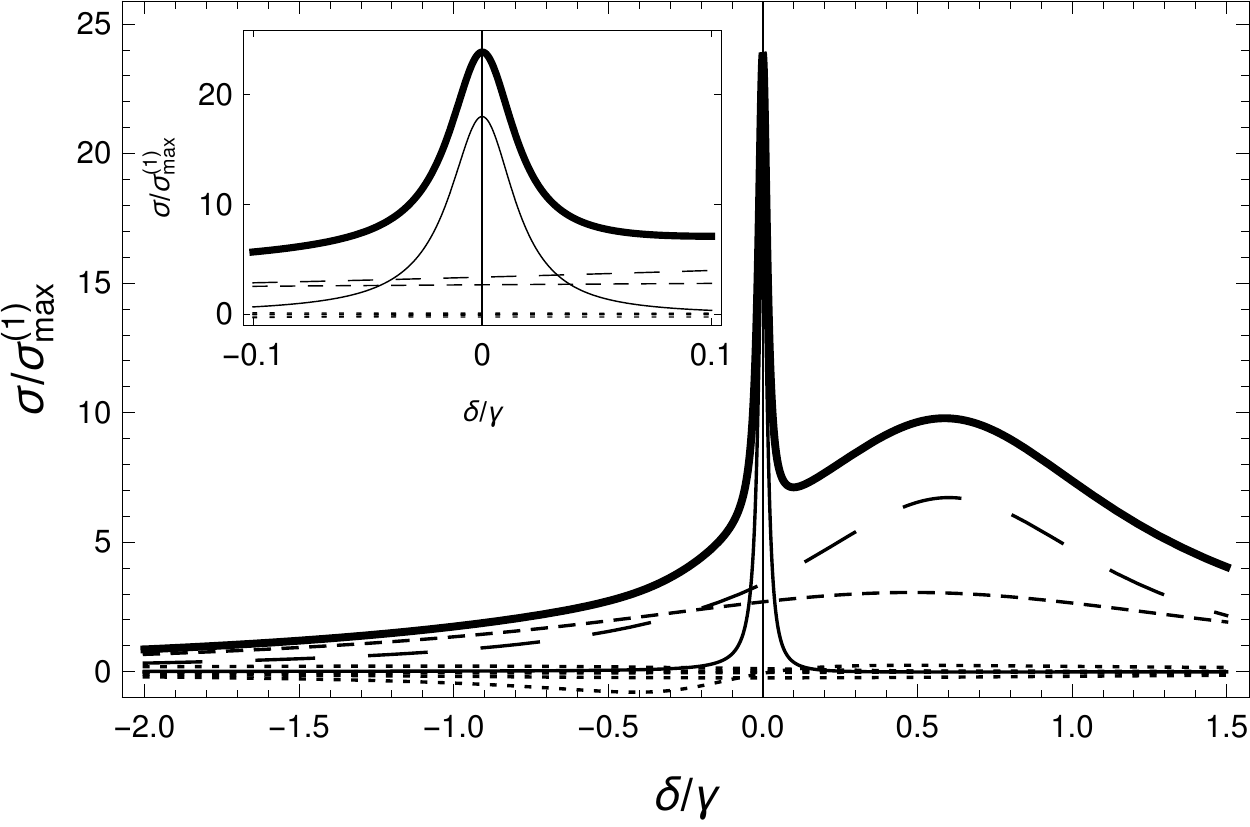}
	\end{center}
	\caption{Optimized scattering cross sections $\sigma$ for eight scatterers in the wide configuration (see figure \ref{fig:config_breit}) in units of the single-scatterer cross section $\sigma^{(1)}_{\rm max}=4\pi/k^2$ as a function of the detuning $\delta/\gamma$ (cf. eq. (\ref{eq:fdelta})). The contributions from the eight different eigenstates as well as the total scattering cross section (thick solid line), i.e. the sum over all scattering resonances, are shown separately. Mainly one single, narrow scattering resonance (thin solid line) contributes to the optimal scattering cross section at $\delta=0$.
	Two further, broad resonances (long-dashed and dashed line) give less important contributions, whereas the remaining five resonances (dotted lines) are negligibly small. The inset shows a zoom around the $\delta=0$ region.
	\label{fig:delta_8_breit}}	
\end{figure}

In this section, we will further analyze the properties of the optimized configurations in terms of their scattering resonances.
As an example, figure \ref{fig:delta_8_breit} shows the scattering cross sections $\sigma_n(\delta)$ associated to the individual resonances, as well as the total scattering cross section $\sigma(\delta)$, see Eq.~(\ref{eq:sigman}), for the optimal wide configuration of $N=8$ scatterers as a function of the detuning $\delta$ of the incoming wave's frequency from the scatterers' resonance frequency. Since the configuration is optimized for the case $\delta=0$, it is not surprising that the total scattering cross section exhibits a sharp maximum around $\delta=0$.
This maximum arises from a single, narrow resonance that gives the main contribution to the total scattering cross section at $\delta=0$.
As we have checked, the same is true also for the optimal narrow configuration (where the scattering cross sections are similar to the ones shown in Fig.~\ref{fig:delta_8_breit}), and also for other numbers $N$ of scatterers.
We therefore conclude that the maximization of the scattering cross section essentially relies on the formation of one single narrow scattering resonance, rather than on finetuning the interplay between several resonances.    

\subsection{Minimization of the smallest decay rate}
\label{sec:mindecay}
\begin{figure}\begin{center}\includegraphics[width=0.9\textwidth]{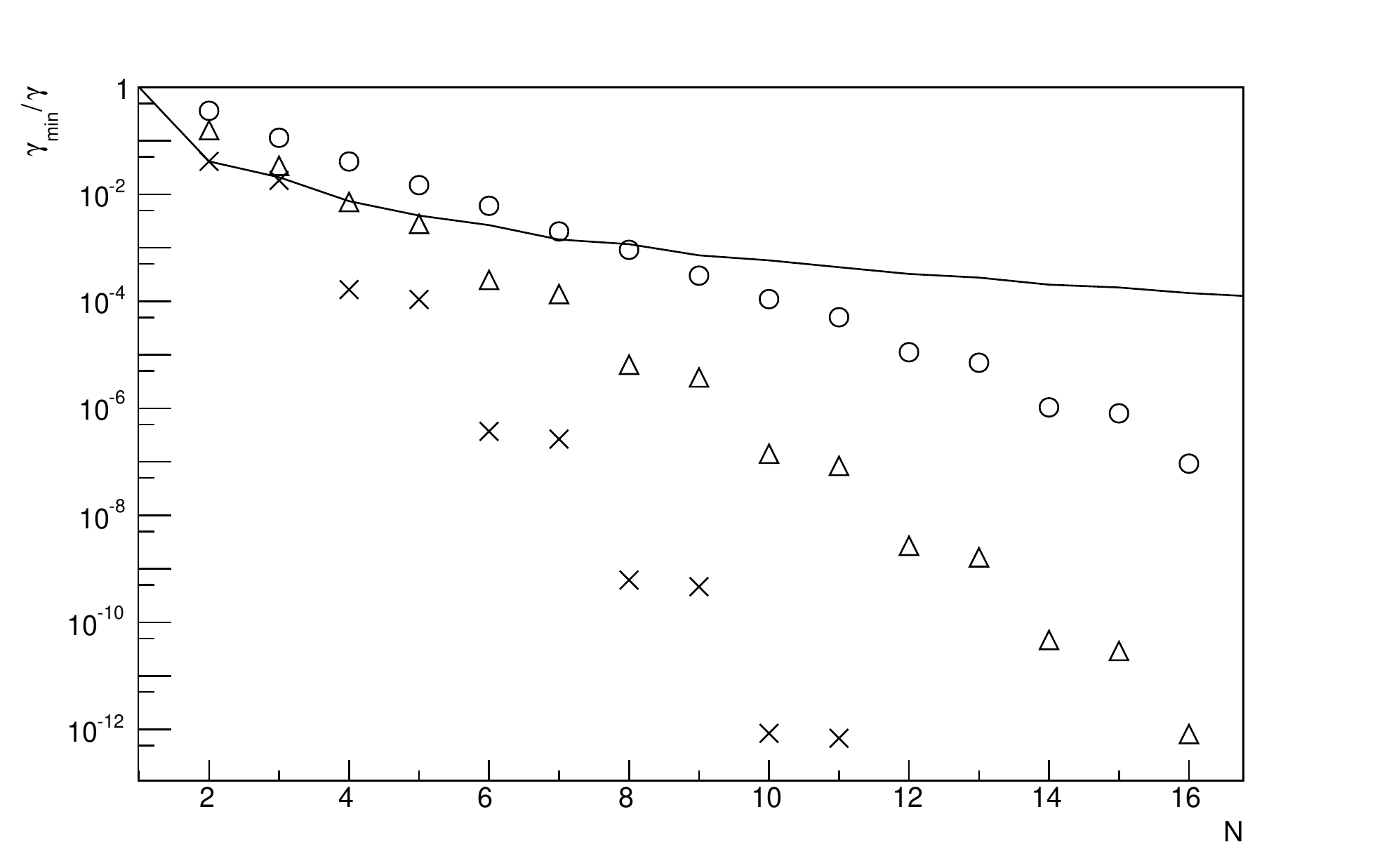}
\caption{Minimal decay rate $\gamma_{\rm min}$ (in units of the decay rate $\gamma$ of a single scatterer) as a function of the number $N$ of resonant point scatterers, for different values of the exclusion radius $kr_{\rm excl}=0.5$ (crosses), $1$ (triangles) and $\pi/2$ (circles).
The minimal decay rate decreases exponentially with $N$. For the smallest value $kr_{\rm excl}=0.5$ of the exclusion radius (crosses), the decay rate is suppressed by 12 orders of magnitude using only $N=10$ scatterers. The solid line shows the result for equally spaced scatterers (with distance $r_{\rm excl} = 0.5k^{-1}$), where the minimal decay rate 
scales approximately like $N^{-3}$.
\label{fig:minimal_decay}}
\end{center}
\end{figure}

\begin{figure}\begin{center}\includegraphics[width=0.9\textwidth]{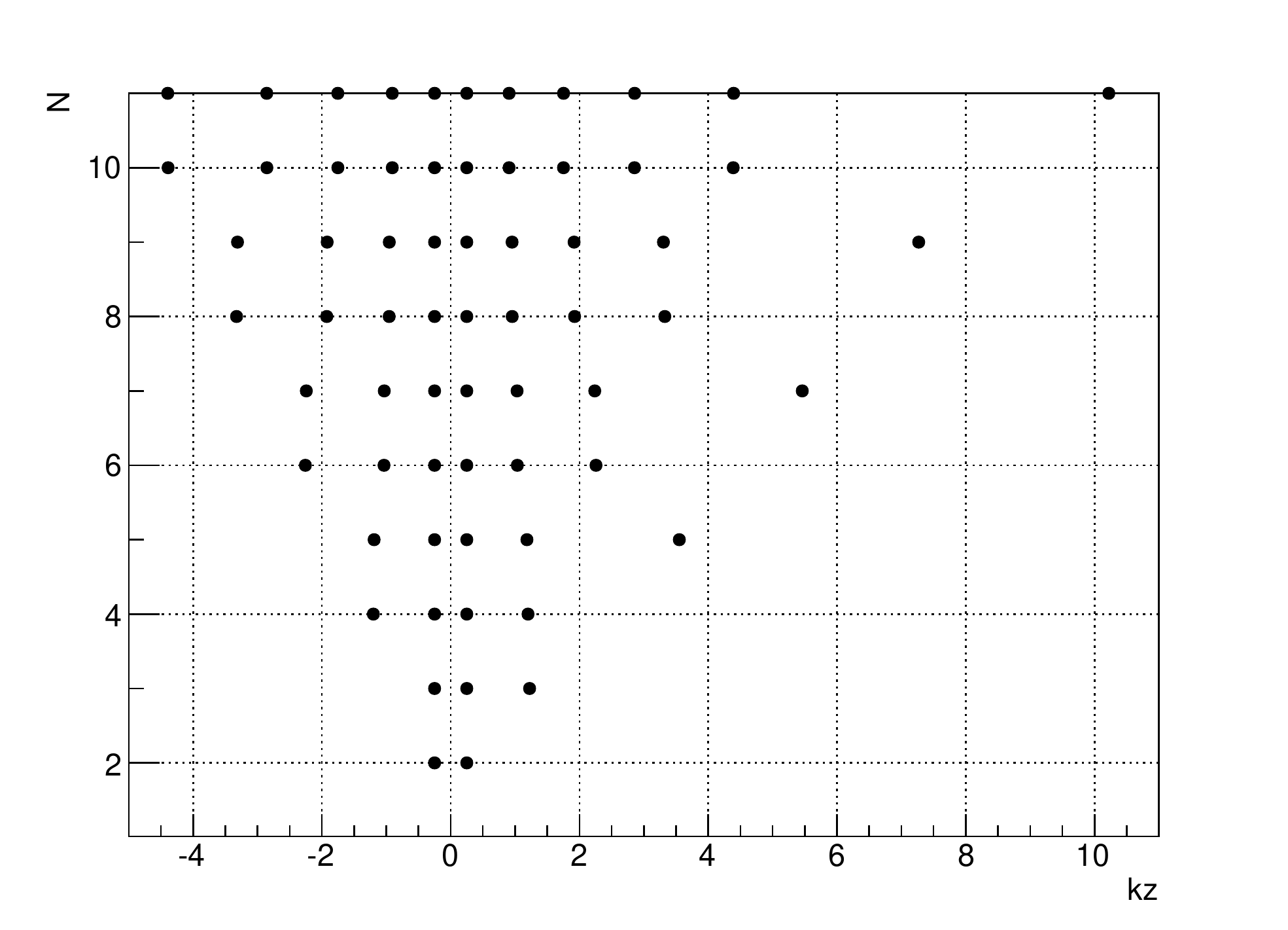}
\caption{Optimized configurations of resonant point scatterers with exclusion radius $r_{\rm excl}=0.5k^{-1}$ exhibiting an eigenstate with minimal decay rate $\gamma_{\rm min}$. As in figures~\ref{fig:config_eng} and \ref{fig:config_breit}, all scatterers are placed on a line, and
configurations with even number $N$  of scatterers are mirror-symmetric with respect to $z=0$. 
The distance between the innermost scatterers equals $r_{\rm excl}$ for all $N$.
\label{fig:config_decay}}
\end{center}
\end{figure}

In the previous section, we have investigated the total scattering cross section resulting from the sum of all $N$ scattering resonances, which are associated to the $N$ eigenvalues of the Green matrix, see chapter~\ref{sec:resonances}. In this section, we are interested in minimizing the decay rate resulting from the smallest imaginary part among these $N$ eigenvalues, see equation~(\ref{eq:gamman}). The result of our numerical optimization is shown in figure~\ref{fig:minimal_decay}, for different values of the exclusion radius $r_{\rm excl}$ defining the smallest possible distance between two scatterers. We see that the minimal decay rate decreases exponentially with $N$ for all three values of $r_{\rm excl}$, all of which fulfill the condition $r_{\rm excl}<\pi k^{-1}$ (see also the discussion at the end of this section). This decrease is faster for smaller values of  $r_{\rm excl}$. 
For the smallest value $r_{\rm excl}=0.5~k^{-1}$ of the exclusion radius (crosses), the decay rate is suppressed by 12 orders of magnitude using only $N=10$ scatterers. In contrast, for equally spaced scatterers with distance smaller than $\pi k^{-1}$ (e.g., $r_{\rm excl}=0.5~k^{-1}$, see the solid line in Fig.~\ref{fig:minimal_decay}), the minimal decay rate decreases only algebraically as a function of $N$ (with exponent $-3$ \cite{felix,tsoi}).

The corresponding optimized configurations, see figure~\ref{fig:config_decay}, display similar properties as those that maximize the scattering cross section:
The configurations that achieve the lowest decay rates are always one-dimensional configurations with an increasing spacing between the scatterers from the center towards the outer edge. While the scatterers are arranged symmetrically for even $N$, the optimal configuration for odd numbers of scatterers is realised by placing an additional scatterer well outside the exclusion radius of the others. As a result, we obtain in figure~\ref{fig:minimal_decay} (crosses) again a pattern originating from the change between even and odd numbers of scatterers, which is more pronounced than  for the  scattering cross section, compare with figure~\ref{fig:crosssec}. In spite of the extremely long lifetimes of the optimized scattering resonances, the associated scattering cross sections are rather small, see, e.g., the values indicated in table \ref{tab:appendixB} in \ref{sec:appendixB} for $N=8$ and $N=9$. As already discussed at the end of section~\ref{sec:decayrates}, this is due to the fact that it is difficult to excite the long-lived resonance  by an incoming plane wave.

Finally, let us shortly comment on the case where the distances between the atoms exceed the value $\pi k^{-1}$. This case is not included in Fig.~\ref{fig:minimal_decay}, since, as we have found, the suppression of the minimal decay rate due to collective effects then becomes considerably less efficient: both for equally spaced scatterers \cite{felix} and for optimized configurations, our numerical investigations indicate that, in this case, the minimal decay rate is always bounded between $\gamma/2$ and $\gamma$ for all values of $N$, and, in particular, does not decrease to zero in the limit $N\to\infty$.

\section{Stability with respect to small deviations from the optimized positions}
\label{sec:stability}

\begin{figure}\begin{center}\includegraphics[width=0.9\textwidth]{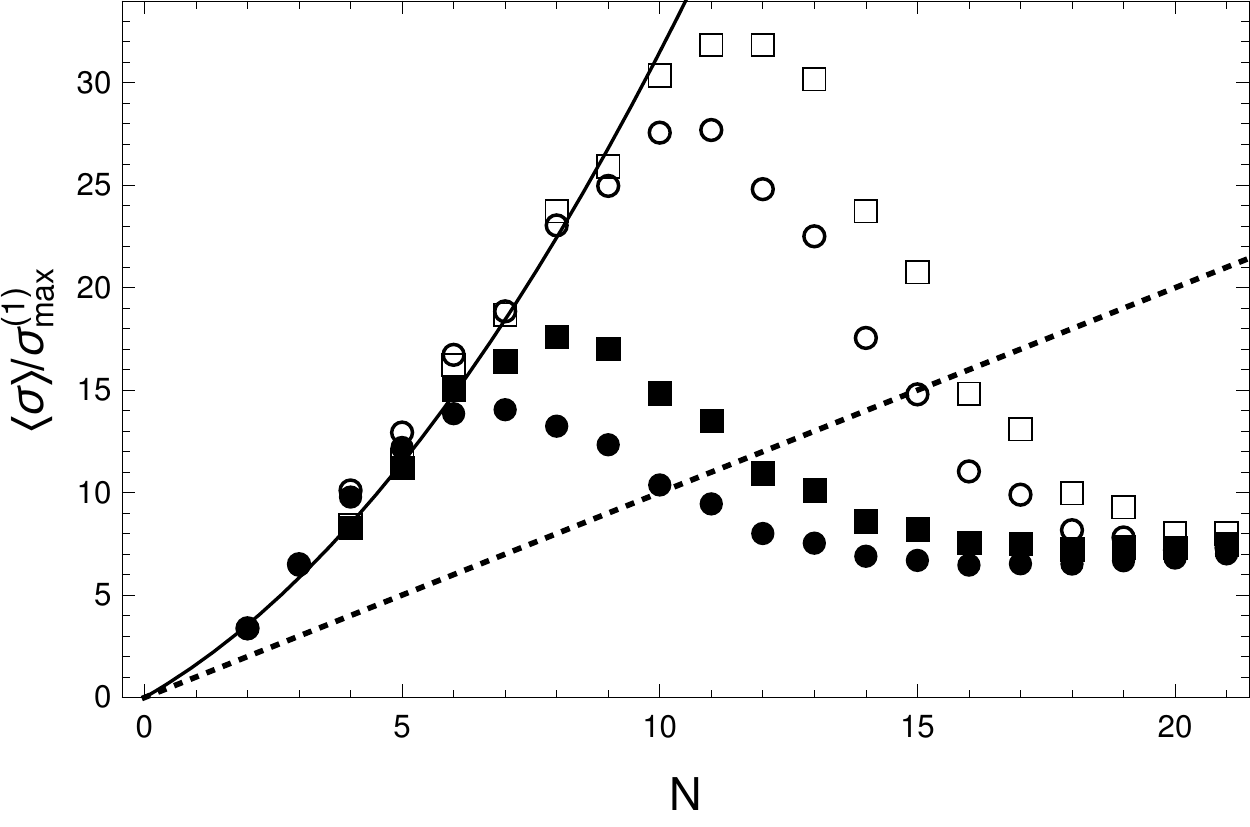}
\caption{Average scattering cross section $\langle\sigma\rangle$ (in units of $\sigma^{(1)}_{\rm max}=4\pi/k^2$) in the presence of small fluctuations of size $\delta r=0.005 k^{-1}$ (open symbols) and 
$\delta r=0.05 k^{-1}$ (filled symbols) around the optimal positions. The circles refer to the 
narrow optimized configurations shown in figure~\ref{fig:config_eng}, whereas the squares refer to the 
wide optimized configurations shown in figure~\ref{fig:config_breit}. 
For small $N$, the average scattering cross sections for both values of the fluctuation radius lie almost on top of each other, and almost coincide with the optimized scattering cross section, cf. figure~\ref{fig:crosssec}.  With increasing $N$, however, the scattering cross sections are increasingly sensitive to fluctuations of the scatterers' positions, where the wide configurations (squares) are more robust than the narrow ones (circles). 
The average scattering cross section exhibits a pronounced maximum as a function of $N$. For comparison, the dotted line displays the scattering cross section for independent scatterers, which are placed far away from each other. The solid line represents the fit to the optimized scattering crossed sections, see Fig.~\ref{fig:crosssec}.
\label{fig:stability_crosssec}}
\end{center}
\end{figure}

\begin{figure}\begin{center}\includegraphics[width=0.9\textwidth]{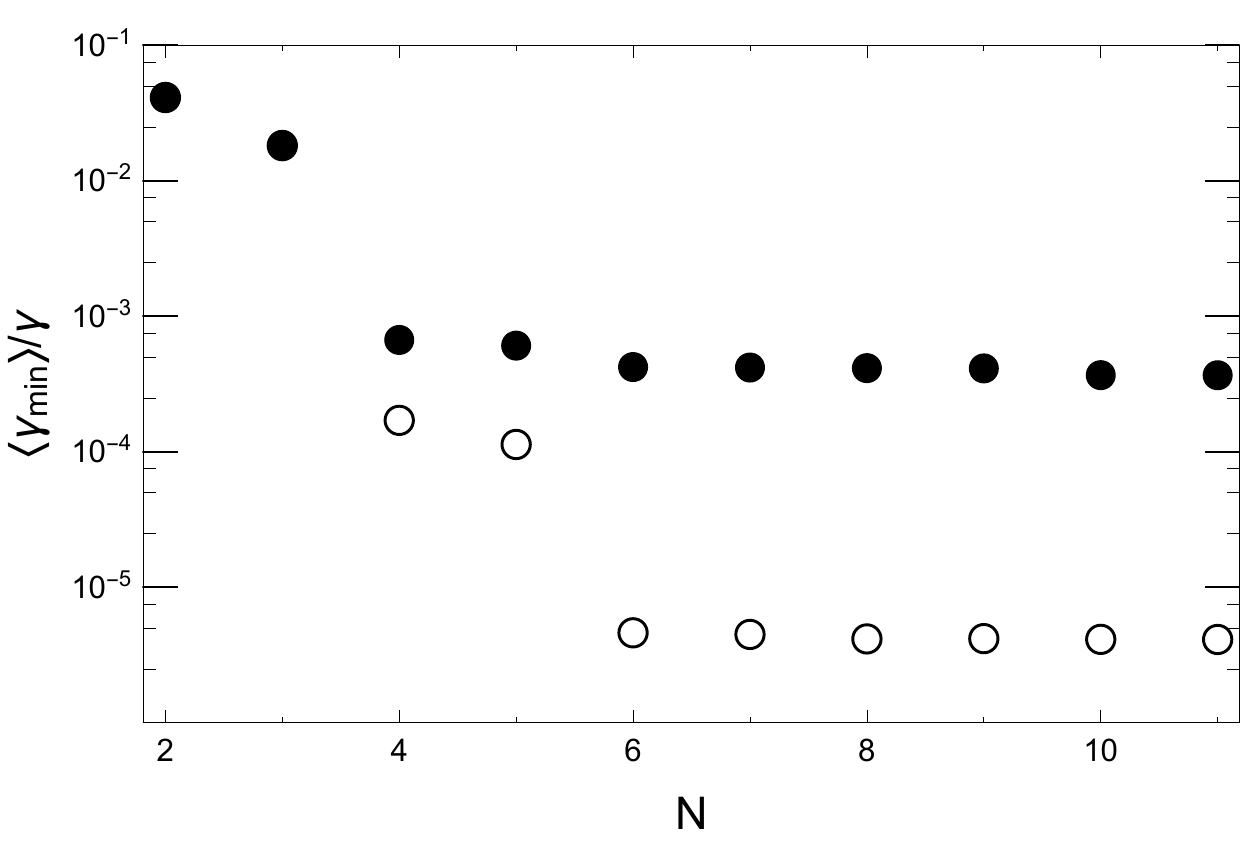}
\caption{Average smallest decay rate $\langle\gamma_{\rm min}\rangle$ (in units of $\gamma$) in the presence of small fluctuations of size $\delta r=0.005 k^{-1}$ (open circles) and 
$\delta r=0.05 k^{-1}$ (closed circles) around the optimal positions for $r_{\rm excl}=0.5k^{-1}$ shown in figure~\ref{fig:config_decay}. $\langle\gamma_{\rm min}\rangle$ saturates as a function of $N$ at a value which is determined by the size of the fluctuations.
Nevertheless, for  $\delta r=0.05 k^{-1}$ (closed circles), the decay rate is suppressed by e.g. more than three orders of magnitude using $N=4$ scatterers, and by more than five orders of magnitude  for  $\delta r=0.005 k^{-1}$ (open circles) using $N=6$ scatterers.
\label{fig:stability_decay}}
\end{center}
\end{figure}

In chapters~\ref{sec:maxcrosssec} and \ref{sec:mindecay}, we have seen that the maximal scattering cross section grows quadratically and the minimal decay rate decreases exponentially with the number $N$ of scatterers -- provided that, in each case, the positions of all scatterers are chosen in an optimal way. From a practical point of view, it is important to know how accurately the 
optimized configurations must be realized, or, in other words, how stable the results are if the positions of the scatterers are allowed to fluctuate around their optimized values.

To answer this question, we start from the corresponding optimized configurations displayed in
figures~\ref{fig:config_eng}, \ref{fig:config_breit} and \ref{fig:config_decay}, and allow the position of each scatterer to deviate from its optimal position by at most $\delta r=0.005 k^{-1}$ (filled symbols) and $\delta r=0.05 k^{-1}$ (open symbols). 
More precisely, each position is chosen randomly according to a uniform  distribution  inside a three-dimensional sphere with radius $\delta r$ surrounding the optimal position. Then, we determine the average scattering cross section or the average smallest decay rate, respectively, where the average is taken over $10^5$ random configurations.

In figure~\ref{fig:stability_crosssec}, we see that, with increasing number $N$ of scatterers, the optimized configurations are increasingly sensitive against small random fluctuations. This applies especially to the narrow configurations (circles), which are less robust than  the wide ones (squares). Furthermore, it is evident that, for larger values of $N$,
the configurations which have been optimized under the assumption that the positions of the scatterers can be precisely controlled, are no longer optimal when taking into account small fluctuations: after reaching a maximum, the average scattering cross section decreases as a function of $N$ and even drops below the value reached by independent scatterers (dotted line).

Fig.~\ref{fig:stability_decay} shows the corresponding results for the smallest decay rate, for fluctuations around the configurations depicted in figure~\ref{fig:config_decay} (with exclusion radius $r_{\rm excl}=0.5 k^{-1}$). The average smallest decay rate saturates as a function of $N$ at a value which is determined by the size of the fluctuations. Remarkably, even in the presence of fluctuations with size $\delta r=0.05 k^{-1}$ ($\delta r=0.005 k^{-1}$), it is possible to suppress the decay rate by more than three (five) orders of magnitude below the decay rate $\gamma$ of a single scatterer.

\section{Conclusion \& Outlook}

We investigated cooperative effects of multiple scattering of scalar waves by finitely many resonant point scatterers 
in 3D. In particular, we found the maximal scattering cross section as well as the minimal decay rate by numerically optimizing the positions  of the individual scatterers. In both cases, the optimum is realized by configurations where the scatterers are positioned on a line parallel to the direction of the incoming light. Exploiting multiple scattering, we consequently could achieve a quadratic increase of the scattering cross section as well as an exponential decrease of the minimal decay rate (in the presence of an exclusion radius $r_{\rm excl}<\pi k^{-1}$ limiting the smallest distance between two atoms) with increasing number of scatterers.  
We showed that the existence of a narrow scattering resonance is a necessary, but not sufficient condition for achieving a large scattering cross section. In particular, the maximization of the total scattering cross section goes along with a concentration of its angular profile into the backward and the forward direction, whereas scattering into all other directions is kept as small as possible.
Allowing for experimental imperfections in the precise realization of the optimized configurations, we also
investigated their stability with respect to small fluctuations of the scatterers' positions.

Since many experiments on multiple scattering are performed with light waves, an obvious extension of our work will be to repeat our analysis for the case of vectorial waves. We expect that the optimal configurations for vectorial scatterers  display similar properties as for scalar waves and, in particular, that the scattering cross section (the smallest decay rate) can be significantly enhanced (reduced) by choosing optimized configurations as compared to regular arrangements with equal spacing between the scatterers. With a view at the experiments on Bragg reflection of light by  atoms mentioned in the introduction \cite{Corzo,Sorensen}, such optimized configurations could be used to achieve even stronger reflection with even smaller numbers of atoms, provided that the positions of the atoms can be controlled with the necessary precision.

Another potential application of our work is the realization of strongly subradiant states, which, due to their long lifetimes, could be good candidates for new types of quantum memories. As we have shown, the decay rate can be suppressed by several orders of magnitude by choosing suitable configurations of point-like emitters. 
However, we also pointed out that it is difficult to excite these long-lived states by an incident wave. Possible remedies are to use a non-optical way of excitation, or to shift the positions of the emitters after they have been excited. Finally, it will be an interesting, but also very challenging task, to extend our studies to the quantum mechanical regime of multiple excitations.

\appendix

\section{An attempt to derive an upper bound of the scattering cross section}
\label{sec:appendix}

We first calculate the decay rate of a scattering resonance $|{\bf \lambda_n^{(R)}}\rangle$, which is defined as right-eigenvector of $G$, i.e. $G|{\bf \lambda_n^{(R)}}\rangle=\lambda_n|{\bf \lambda_n^{(R)}}\rangle$
Writing $|{\bf \lambda_n^{(R)}}\rangle=\sum_i c^{(n)}_i |i\rangle$ with $\sum_i |c^{(n)}_i|^2=1$ (i.e. $\langle{\bf \lambda^{(R)}_n}|{\bf\lambda^{(R)}_n}\rangle=1$), we obtain, see Eq.~(\ref{eq:gamman}):
\begin{eqnarray}
\gamma_n & = & \gamma\left(1+{\rm Im}\left[\langle{\bf \lambda^{(R)}_n}|G|{\bf\lambda^{(R)}_n}\rangle\right]\right)\nonumber\\
& = & \gamma\left(1+\sum_{i\neq j} c^{(n)}_i c^{(n)*}_j \frac{\sin(kr_{ij})}{kr_{ij}}\right)\nonumber\\
& = & \gamma \int\frac{{\rm d}\Omega}{4\pi}  \left|\sum_{i=1}^N c^{(n)}_ie^{i{\bf r}_i\cdot{\bf k}_\Omega}\right|^2\label{eq:gammatheo}
\end{eqnarray}
Here, $\int{\rm d}\Omega$ denotes an integral over the angular variables of ${\bf k}_\Omega$ (with $|{\bf k}_\Omega|=k$).
We used:
\begin{equation}
\frac{\sin(k r_{ij})}{kr_{ij}}= \frac{1}{2}\int_{-1}^1 {\rm d}\cos(\theta)~ e^{i\cos(\theta) k r_{ij}}= \int\frac{{\rm d}\Omega}{4\pi} e^{i({\bf r}_i-{\bf r}_j)\cdot{\bf k}_\Omega}\label{eq:OmG}
\end{equation}
to arrive at the third line of (\ref{eq:gammatheo}). We define
\begin{equation}
g_n({\bf k})=\sum_{i=1}^N c^{(n)}_i e^{-i{\bf r}_i\cdot{\bf k}}
\label{eq:gk}
\end{equation}
We note that $g_n({\bf k})$ (with $|{\bf k}|=k$) is proportional to the  wave amplitude emitted from the $n$-th resonance into the direction ${\bf k}$, see also (\ref{eq:Psisc}). According to equation (\ref{eq:gammatheo}), the decay rate $\gamma_n$ is determined by the total flux emitted into all directions.

In order to derive an upper bound for $\sigma_n(\delta)$, see equation (\ref{eq:resonancen}), we replace the imaginary part by the absolute value. Then, the maximum achieved at $\delta=\delta_n$ reads:
\begin{equation}
\sigma_n\leq \frac{4\pi\gamma}{k^2\gamma_n|\Psi_0|^2} \left|\langle{\bf \Psi_0}|{\bf \lambda_n^{(R)}}\rangle\langle{\bf \lambda_n^{(L)}}|{\bf \Psi_0}\rangle\right|
\end{equation}
Since the matrix $G$ is complex symmetric (i.e. $G_{ij}=G_{ji}$), we can choose the left eigenvector as the complex conjugate of the right eigenvector, i.e. $\langle {\bf\lambda_n^{(L)}}|=\sum_i c^{(n)}_i \langle i|$. Thereby, we obtain:
$$\frac{\langle{\bf \Psi_0}|{\bf \lambda_n^{(R)}}\rangle\langle{\bf \lambda_n^{(L)}|{\bf\Psi_0}\rangle}}{|\Psi_0|^2}=
\sum_{ij}c^{(n)}_i c^{(n)}_j e^{i{\bf k}_{\rm in}\cdot({\bf r}_i-{\bf r}_j)}=g_n({\bf k}_{\rm in}) g_n(-{\bf k}_{\rm in})$$
This shows that the scattering cross section is invariant with respect to changing the direction of the incoming wave (${\bf k}_{\rm in}\to -{\bf k}_{\rm in}$).
The normalization condition
$\langle {\bf\lambda_n^{(L)}}|{\bf \lambda_n^{(R)}}\rangle=1$ reads $\sum_i (c^{(n)}_i)^2 =1$, which differs from the above
condition $\sum_i |c^{(n)}_i|^2 =1$ for $\langle {\bf\lambda_n^{(R)}}|{\bf \lambda_n^{(R)}}\rangle=1$. Taking this into account, we finally obtain:
\begin{equation}
\frac{\sigma_n}{\sigma^{(1)}_{\rm max}} \leq \left|\frac{g_n({\bf k}_{\rm in}) g_n(-{\bf k}_{\rm in})}{\sum_i (c^{(n)}_i)^2}\right|
\frac{\sum_i |c^{(n)}_i|^2}{\int\frac{{\rm d}\Omega}{4\pi} |g_n({\bf k}_\Omega)|^2}
\label{eq:upperbound}
\end{equation}
The upper bound is saturated if and only if the expression inside the absolute value is real and positive, i.e. if $g_n({\bf k}_{\rm in})g_n(-{\bf k}_{\rm in})/\sum_i (c^{(n)}_i)^2>0$.

This result clarifies the qualitative discussion in chapter~\ref{sec:decayrates}. In order to maximize the scattering cross section,
one has to achieve large emissions $|g_n({\bf k}_{\rm in})|$ and  $|g_n(-{\bf k}_{\rm in})|$ into the forward and backward directions $\pm{\bf k}_{\rm in}$ while keeping the emission into all other directions as small as possible. At the same time, the amplitudes $c_i$ of the resonant state should be optimized such that $\left|\sum_i (c^{(n)}_i)^2\right|$ is minimized under the constraint
$\sum_i |c^{(n)}_i|^2=1$. 

The latter expression is, in principle, not bounded from below. This causes the main difficulty in deriving an upper bound of the scattering cross section. Indeed, it may even happen that $\left|\sum_i (c^{(n)}_i)^2\right|=0$ for certain configurations. This is the case if the Green matrix $G$ is not diagonalisable. An example of such a case for $N=3$ is obtained if
the distances between the three scatterers are chosen as $k r_{12}=k r_{13} = 3 \pi (4+\sqrt{2})/7$ and $k r_{23}=3\pi (1+2\sqrt{2})/14$ (which corresponds to a two-dimensional configuration). In this case, contributions from invidual scattering resonances give a diverging result, although, as we have checked, the total scattering cross section defined by (\ref{eq:crosssectionvec}) remains finite (and smaller than the scattering cross section of the optimized configuration).

\section{Optimized configurations for $N=8$ and $N=9$}
\label{sec:appendixB}

In Table~\ref{tab:appendixB}, we give the positions of all scatterers in the optimized configurations for $N=8$ and $N=9$, together with their total cross sections and minimal decay rates.

\begin{table}[t]
\begin{center}	
	\begin{tabular}{r c p{7.5cm}cc}
	\toprule
  	$N$ & type & $kz_1,..,kz_N$ & $\sigma$/$\sigma^{(1)}_{\rm max} $ & $\gamma_{min}/\gamma$ \\
	\midrule
 	8 & narrow & -8.341, -5.411, -2.908, -0.830, \newline 0.830, 2.908, 5.411, 8.341   &  23.6 &  $6.8\cdot 10^{-3}$ \\
 	
 	8 & wide & -8.452, -5.667, -3.453, -2.004,   \newline 2.004, 3.453, 5.667, 8.452    &  23.9 &  $1.7\cdot 10^{-2}$ \\
 	
 	8  & decay &  -3.32587, -1.92458, -0.95543, -0.25, \newline 0.25, 0.95543, 1.92458, 3.32587  &  1.3 &  $6.2\cdot 10^{-10}$ \\

 	& & & & \\
 	
 	9 & narrow & -8.340, -5.458, -2.960, -0.843, 0.843,\newline 2.851, 5.273, 8.100,  11.302 &  26.1 &  $4.4\cdot 10^{-3}$ \\
 	
 	9 & wide &  -8.423, -5.679, -3.444, -1.989,  1.989, \newline 3.460, 5.599, 8.301, 11.479  &  26.2 &  $1.2\cdot 10^{-2}$ \\
 	
 	9 & decay & -3.31104, -1.91837, -0.95357, -0.25, 0.25, \newline 0.95338, 1.91681, 3.30431, 7.26926 &  1.7 &  $4.6\cdot 10^{-10}$\\
 	\bottomrule
	\end{tabular}
\end{center}
\caption{Positions in units of $k^{-1}$, scattering cross sections $\sigma$ in units of the single scatterer cross section $\sigma^{(1)}_{\rm max}=4\pi/k^2$ and the smallest decay rates $\gamma_{min}$ in units of the decay rate $\gamma$ of a single scatterer of the optimized configurations for eight and nine scatterers. The three different types of configurations (narrow, wide and decay) correspond to the ones shown in Fig. \ref{fig:config_eng} (narrow), Fig. \ref{fig:config_breit} (wide) and Fig. \ref{fig:config_decay} (decay).\label{tab:appendixB}}
\end{table}			



\section*{References}

\begin{thebibliography}{9}
	
	\bibitem{Bragg}Bragg W H and Bragg W L 1913 The Reflection of X-rays by Crystals, {\it Proc R. Soc. Lond. A.} \textbf{88}, 428
	
	\bibitem{Corzo}Corzo N V, Gouraud B, Chandra A, Goban A, Sheremet A S, Kupriyanov D V, and Laurat J 2016 Large Bragg Reflection from One-Dimensional Chains of Trapped Atoms Near a Nanoscale Waveguide, {\it Phys. Rev. Lett.} \textbf{117}, 133603 

	
	\bibitem{Sorensen}S\o rensen H L, B\'eguin J-B, Kluge K W, Iakoupov I, S\o rensen A S, M\"uller J H, Polzik E S, and Appel J 2016 Coherent Backscattering of Light Off One-Dimensional Atomic Strings, {\it Phys. Rev. Lett.} \textbf{117}, 133604	
			
	\bibitem{hofmann} Hofmann C L M, Herter B, Fischer S, Gutmann J and Goldschmidt J C 2016  Upconversion in a Bragg structure: photonic effects of a modified local density of states and irradiance on luminescence and upconversion quantum yield, Opt. Expr. {\textbf 24} 14895
	
	\bibitem{spallek} Spallek F, Buchleitner A, and Wellens T 2017 Optimal trapping of monochromatic light in designed photonic multilayer structures, arXiv:1706.05079

	\bibitem{podolskyi} Podolskiy V, Sarychev A, Narimanov E and  Shalaev V 2005 Resonant light interaction with plasmonic
		nanowire systems, {\it J. Opt. A: Pure Appl. Opt.} \textbf{7}  S32
	
	\bibitem{Li} Li K, Stockman M I, and Bergman D J 2003 Self-similar chain of metal nanospheres as an efficient nanolens, 
	{\it Phys. Rev. Lett.} \textbf{91}, 227402    
	
	
	\bibitem{Wang} Wang Z B, Luk’yanchuk B S, Guo W, Edwardson S P,  Whitehead D J, Li L, Liu Z and Watkins K G 2008 The influences of particle number on hot spots in strongly coupled metal nanoparticles chain, {\it J. Chem. Phys.}
	\textbf{128}, 094705 
	
	\bibitem{niuwenhuizen} Niuwenhuizen T M, Lagendijk A, and van Tiggelen B A 1992 Resonant point-scatterers in multiple scattering of classical waves, {\it Phys. Lett. A} \textbf{169}, 191
	
	\bibitem{pavolini} Pavolini D, Crubellier A, Pillet P, Cabaret L and Liberman S 1985 Experimental Evidence for Subradiance {\it Phys. Rev. Lett.} \textbf{54} 1917
	
	\bibitem{guerin} Guerin W, Ara\'ujo M O, and Kaiser R 2016 Subradiance in a Large Cloud of Cold Atoms {\it Phys. Rev. Lett.} \textbf{116}, 083601
		
	\bibitem{foldy} Foldy L L 1945 The Multiple Scattering of Waves. I. General Theory of
		Isotropic Scattering by Randomly Distributed Scatterers, {\it Phys. Rev.} \textbf{67}, 107
	
	\bibitem{lax} Lax M 1951 Multiple Scattering of Waves, {\it Rev. Mod. Phys.}
	\textbf{23}, 287
	
	
	
	\bibitem{rossum} van Rossum N C W and Nieuwenhuizen T M 1999 Multiple scattering of classical waves: microscopy, mesoscopy, and diffusion {\it Rev. Mod. Phys.} \textbf{71} 313
	
	\bibitem{skipetrov} Skipetrov S E and Sokolov I M 2014 Absence of Anderson Localization of Light in a Random Ensemble of Point Scatterers {\it Phys. Rev. Lett.} \textbf{112} 023905

	\bibitem{bettles} Bettles R J, Gardiner S A and Adams C S 2016 Cooperative eigenmodes and scattering in one-dimensional atomic arrays {\it Phys. Rev. A} \textbf{94} 043844 
	
	\bibitem{lippmann} Lippmann B A and Schwinger J Variational Principles for Scattering Processes. I 1950 {\it Phys. Rev.} \textbf{79} 469 
		

	
	\bibitem{heller1} Li S and Heller E J 2003 Quantum multiple scattering: Eigenmode
		expansion and its applications to proximity resonance, {\it Phys. Rev. A} \textbf{67}, 032712
	
	\bibitem{rodberg} Rodberg L S and Thaler R M 1967 \emph{Introduction to the quantum theory of
		scattering}, Academic Press, New York, London

	
	\bibitem{fano}Fano U 1961 Effects of Configuration Interaction on Intensities and Phase Shifts, {\it Phys. Rev.} \textbf{124}, 1866
	
	\bibitem{heller2} Heller E J 1996 Quantum Proximity Resonances, {\it Phys. Rev. Lett.} \textbf{77}, 4122 
	
	
	\bibitem{c} Press W H, Teukolsky S A, Vetterling W T and Flannery B P 2002 \emph{Numerical Recipes in C}, 2nd Edition, Cambridge University Press
	
	\bibitem{frank} Sch\"afer F 2015 \emph{Optimal configurations for linear point scatterers}, Bachelor Thesis, Albert-Ludwigs-Universit\"at Freiburg
	
	\bibitem{felix} Eckert F 2013 \emph{Speckle Instabilities in non-linear disordered media}, Dissertation, Albert-Ludwigs-Universit\"at Freiburg
	
	\bibitem{tsoi} Tsoi T S and Law C K 2008 Quantum interference effects of a single photon interacting with an atomic chain inside a one-dimensional waveguide
	{\it Phys. Rev. A} \textbf{78} 063832 
		
\end{thebibliography}
\end{document}